\documentclass[preprint,amssymb,showpacs,superscriptaddress]{revtex4}

\usepackage{graphicx}
\usepackage{dcolumn}
\usepackage{bm}
\usepackage{color}

\begin{document}

\preprint{preprint(\today)}

\title{Tuning the static spin-stripe phase and superconductivity in La$_{2-x}$Ba$_{x}$CuO$_{4}$ ($x$ = 1/8) by hydrostatic pressure}

\author{Z.~Guguchia}
\email{zurabgug@physik.uzh.ch} \affiliation{Physik-Institut der
Universit\"{a}t Z\"{u}rich, Winterthurerstrasse 190, CH-8057
Z\"{u}rich, Switzerland}

\author{A.~Maisuradze}
\affiliation{Physik-Institut der Universit\"{a}t Z\"{u}rich,
Winterthurerstrasse 190, CH-8057 Z\"{u}rich, Switzerland}
\affiliation{Laboratory for Muon Spin Spectroscopy, Paul Scherrer Institute, CH-5232
Villigen PSI, Switzerland}

\author{G.~Ghambashidze}
\affiliation{Department of Physics, Tbilisi State University,
Chavchavadze 3, GE-0128 Tbilisi, Georgia}

\author{R.~Khasanov}
\affiliation{Laboratory for Muon Spin Spectroscopy, Paul Scherrer Institute, CH-5232
Villigen PSI, Switzerland}

\author{A.~Shengelaya}
\affiliation{Department of Physics, Tbilisi State University,
Chavchavadze 3, GE-0128 Tbilisi, Georgia}

\author{H.~Keller}
\affiliation{Physik-Institut der Universit\"{a}t Z\"{u}rich,
Winterthurerstrasse 190, CH-8057 Z\"{u}rich, Switzerland}

\begin{abstract}
 Magnetization and muon spin rotation experiments were performed in La$_{2-x}$Ba$_{x}$CuO$_{4}$ ($x$ = 1/8)
as a function of hydrostatic pressure up to $p$~${\simeq}$~2.2 GPa.
It was found that  the magnetic volume fraction of the static stripe phase strongly decreases linearly with 
pressure, while the superconducting volume fraction 
increases by the same amount. This demonstrates competition 
between bulk superconductivity and static magnetic order in the stripe phase of La$_{1.875}$Ba$_{0.125}$CuO$_{4}$ 
and that these phenomena occur in mutually exclusive spatial regions.
The present results also reveal that the static spin-stripe phase still exists at pressures, where
the long-range low-temperature tetragonal (LTT) structure is completely suppressed. This indicates that
the long-range LTT structure is not necessary for stabilizing the static spin order in La$_{1.875}$Ba$_{0.125}$CuO$_{4}$.

\end{abstract}


\maketitle

La$_{2-x}$Ba$_{x}$CuO$_{4}$ (LBCO) was the first cuprate in which high-$T_{\rm c}$ superconductivity was discovered \cite{Bednorz}.
The undoped parent compound is an antiferromagnetic (AFM) insulator. The replacement of La$^{3+}$ by Ba$^{2+}$ ions, through which holes are doped into the CuO$_{2}$ planes, causes the destruction of 
AFM order and superconductivity appears at $x$ = 0.06. Subsequent investigations showed that there exists a sharp dip in the $T_{\rm c}$-$x$ phase diagram, indicating that bulk superconductivity is greatly suppressed in a narrow range around a particular doping concentration $x$ = 1/8 in LBCO \cite{Moodenbaugh}. 
This  suppression of $T_{\rm c}$ has attracted a great deal of attention and is known in the literature as the 1/8 anomaly (see $e.g.$, \cite{Kivelson,Vojta}). 
Later a similar anomaly was also observed in rare earth doped La$_{2-x}$Sr$_{x}$CuO$_{4}$. 
Studies of the crystal structure clarified that the LBCO system undergoes 
at $x$ = 1/8 a first-order structural phase transition from a low-temperature orthorhombic (LTO) to a low-temperature tetragonal (LTT) phase \cite{Axe}.
Since the structural transition to the LTT phase appears near the Ba concentration $x$ where the strong decrease of $T_{\rm c}$ occurs, it has been suggested that there is a correlation between the appearance of the LTT phase and the suppression of superconductivity \cite{Axe}. Muon spin rotation (${\mu}$SR) experiments detected the appearence of static magnetic order 
below ${\sim}$ 30 K in La$_{1.875}$Ba$_{0.125}$CuO$_{4}$ (LBCO-1/8) \cite{Luke}.

The discovery of elastic superlattice peaks in La$_{1.48}$Nd$_{0.4}$Sr$_{0.12}$CuO$_{4}$ by neutron diffraction 
provided evidence of two-dimensional charge and spin order, which was explained in terms of a stripe model where charge-carrier poor
AFM regions are separated by one-dimensional stripes of charge carrier-rich  regions \cite{Tranquada1,Tranquada2}. The presence of stripe-like charge and spin density ordering is believed to be responsible for the anomalous suppression of superconductivity around $x$ = 1/8 in cuprates \cite{Tranquada1,Tranquada2}. 
The existence of stripes in La$_{1.85}$Sr$_{0.15}$CuO$_{4}$ and Bi$_{2}$Sr$_{2}$CaCu$_{2}$O$_{8+y}$ 
has also been demonstrated by extended x-ray absorption fine structure (EXAFS) 
experiments which allow to probe the local structure near a selected atomic site \cite{Bianconi1,Bianconi2}.

The fascinating issue of charge and spin stripes in cuprate superconductors has attracted a lot of attention for many years (see $e.g.$, \cite{Kivelson,Vojta}). 
Experimental results and theoretical considerations show that the modulations of the lattice and of the charge and spin density appear to be both ubiquitous in the cuprates and intimately tied up with the physics of these materials \cite{Kivelson,Vojta}. However, the role of stripes for superconductivity in cuprates is still unclear at present.
Therefore, it is important to find an external control parameter which allows to tune structural and electronic properties of the cuprates and study the relation between superconductivity and stripe order. It is known that upon applying hydrostatic pressure both the LTT and LTO structural phase transition in LBCO-1/8 are suppressed
completely at the critical pressure $p_{c}$ ${\approx}$ 1.85 GPa, and superconductivity is enhanced \cite{Katano,Crawford,Zhou,Hucker}. The magnetic order related to stripe formation was previously studied under pressure, but only below $p_{c}$ \cite{Satoh}. Hence, it is not known how the static spin-stripe order changes across $p_{c}$.  

 Here, we report studies of superconductivity and stripe magnetic order in LBCO-1/8 under hydrostatic pressure up to $p$ ${\simeq}$ 2.2 GPa  by magnetization and 
muon-spin rotation (${\mu}$SR) experiments. It was observed that the transition temperature of the 
stripe magnetic order and the size of the ordered moment are not significantly changed by pressure. But the
volume fraction of the magnetic phase significantly decreases and simultaneously the superconducting (SC) volume fraction increases with increasing pressure. 
This indicates that magnetic regions in the sample are converted to SC regions 
with increasing pressure, providing evidence for a competition between superconductivity and 
static magnetic order in the stripe phase of LBCO-1/8. 
It was also demonstrated that the spin-stripe order still exists at pressures, where the LTT
phase is suppressed.  
 
One polycrystalline sample of La$_{2-x}$Ba$_{x}$CuO$_{4}$ with $x$ = 1/8 was prepared by 
the conventional solid-state method. All the measurements were performed on samples
from the same batch. The single-phase character of the sample was checked by powder x-ray diffraction.

\begin{figure}[t!]
\centering
\includegraphics[width=0.6\linewidth]{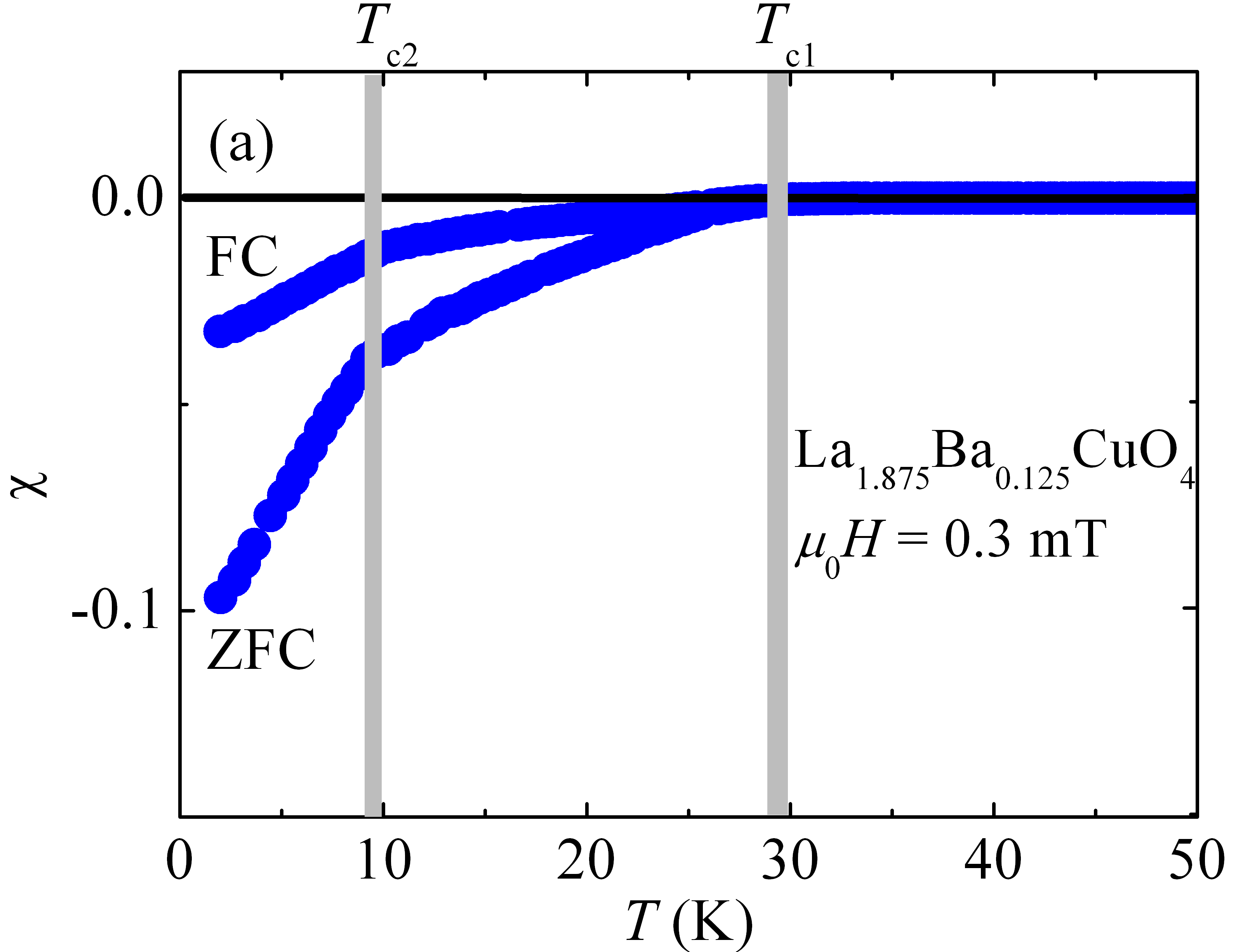}
\includegraphics[width=0.6\linewidth]{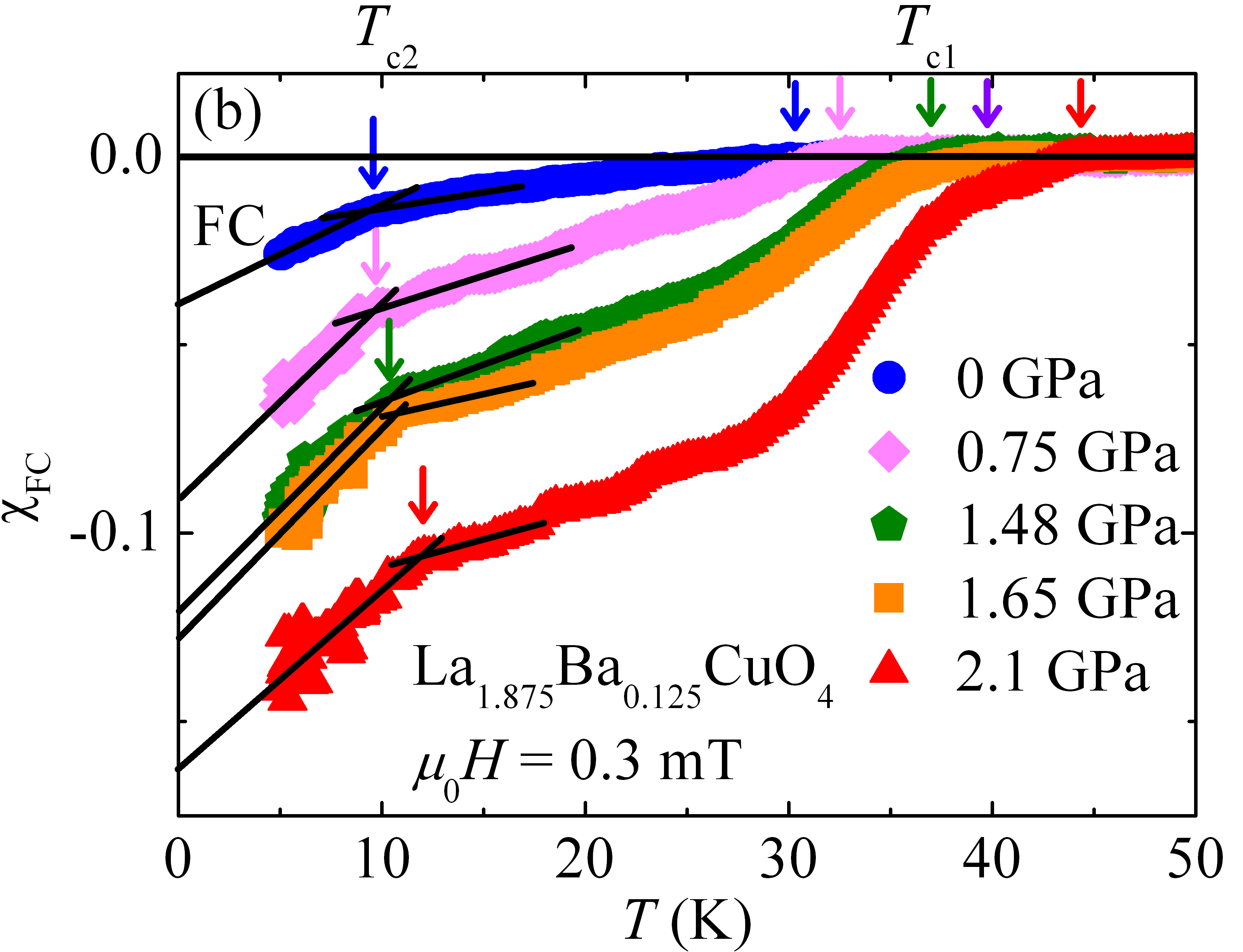}
\vspace{-0.3cm}
\caption{ (Color online) Temperature dependence of the magnetic susceptibility 
of LBCO-1/8
measured at ambient pressure without pressure cell (a) and at various applied hydrostatic pressures (b) in a magnetic field of ${\mu}_{\rm 0}$$H$ = 0.3 mT.
The vertical gray lines and the arrows denote the superconducting transition 
temperatures $T_{\rm c1}$ and $T_{\rm c2}$ (see text for an explanation).}
\label{fig1}
\end{figure}

The magnetic susceptibility was measured under pressures up to 2.1 GPa
by a SQUID magnetometer ($Quantum$ $Design$ MPMS-XL). 
Pressures were generated using a diamond anvil cell (DAC) \cite{Giriat} 
filled with Daphne oil which served as a pressure-transmitting medium. 
The pressure at low temperatures was determined 
by the pressure dependence of the SC transition temperature of Pb.
The temperature dependence of the zero-field-cooled (ZFC) and field-cooled (FC)  
magnetic susceptibility, $\chi_{\rm ZFC}$ and $\chi_{\rm FC}$, respectively, for LBCO-1/8
in a magnetic field of ${\mu}_{\rm 0}$$H$ = 0.3 mT is shown in Fig.~1a.
The diamagnetic susceptibility exhibits a two-step SC transition. 
The first transition with an onset at $T_{\rm c1}$ ${\approx}$ 30 K corresponds to only
about 4 ${\%}$ volume fraction of superconductivity estimated from ZFC magnetization at 10 K.
The second SC transition is observed at $T_{\rm c2}$ ${\approx}$ 10 K, with a 
larger diamagnetic response.  However, the volume fraction of the low temperature SC phase 
is still small at ambient pressure and amounts to about 10 ${\%}$ of full shielding at 2 K.
A two-step SC transition, starting at around 30 K with a weak diamagnetic response
was observed previously in polycrystalline LBCO-1/8 \cite{Katano,Moodenbaugh}.  
It was explained as some kind of filamentary superconductivity due to the presence of a very
small fraction of the LTO phase. Recent detailed transport and susceptibility 
measurements in single crystal of LBCO-1/8 provided evidence of the intrinsic nature 
of the observed two-step SC transition \cite{Tranquada2008}.
It was found that a SC transition at higher temperature $T_{\rm c1}$ is
present when the magnetic field is applied perpendicular to the CuO$_{2}$ planes.
The SC transition at low temperature $T_{\rm c2}$ is more pronounced when the magnetic 
field is parallel to the planes ($H \parallel ab$).
The authors interpreted the transition at $T_{\rm c1}$ as due to 
the development of 2D superconductivity in the CuO$_{2}$ planes, while the interlayer 
Josephson coupling is frustrated by static stripes. A transition to a 3D SC phase
takes place at much lower temperature $T_{\rm c2}$ ${\ll}$ $T_{\rm c1}$, reflected as a strong increase of diamagnetism below $T_{\rm c2}$ 
for $H \parallel ab$. For polycrystalline samples with random orientation of grains these two 
temperatures will result in two SC transitions as observed in present experiments (see Fig.~1a).

We studied the SC transition in LBCO-1/8 as a function of hydrostatic pressure.  
Measurements were performed in the FC mode at 0.3 mT, which was set constant during the measurements at all pressures 
in order to avoid a variation of the applied field during the measurements with different pressures. 
Figure~1b shows the temperature dependence of $\chi_{\rm FC}$ for different pressures after substraction 
of the background signal from the empty pressure cell. A two-step SC transition is observed at all pressures, 
except at the highest applied pressure of 2.1 GPa, where a three-step SC transition is visible.
The reason for this is not clear at present. Further investigations, in particular on single crystals, are needed to clarify this issue.  
It was found that $T_{\rm c2}$ increases only slightly with pressure from 10 K to about 12 K at the maximal pressure applied in our experiments ($p$ = 2.1 GPa). On the other hand, $T_{\rm c1}$ shows a significant increase with a rate of 6.2 K/GPa. It is interesting that the volume fraction of the corresponding SC phase is also strongly enhanced with applied pressure (see Fig.~1b). These results are in agreement with previous studies showing that superconductivity in LBCO-1/8 is largely enhanced by applying 
pressure \cite{Katano,Zhou,Yamada}.

It is interesting to explore the pressure effect on spin order in the stripe phase
and its relation to superconductivity. It is also of great interest to study the relation between static
magnetism and the LTT phase in LBCO-1/8. However, to the best of our knowledge magnetism in LBCO-1/8 was studied only at low pressures \cite{Satoh} where the LTT phase is still present. In order to answer this question we performed zero-field (ZF) ${\mu}$SR experiments in LBCO-1/8 at ambient and various hydrostatic pressures, including pressures where the long-range LTT structure is suppressed. ZF ${\mu}$SR is a powerful tool to investigate microscopic magnetic properties of solids without applying an 
external magnetic field. It is especially suitable for the study of weak magnetic order, since the positive muon 
is an extremely sensitive local probe which is able to detect small internal magnetic fields and ordered volume fractions.

The ZF ${\mu}$SR experiments were carried out at the ${\mu}$E1 beam line at the Paul Scherrer Institute, Switzerland.
Pressures up to 2.2 GPa were generated in a double wall piston-cylinder type of cell made of MP35N material, especially designed to perform ${\mu}$SR experiments under pressure \cite{Maisuradze}. As a pressure transmitting medium Daphne oil was used. 
The pressure was measured by tracking the SC transition of a very small indium plate.
The ${\mu}$SR time spectra were analyzed using the free software package MUSRFIT \cite{Suter}.

Figure~2 shows representative ZF ${\mu}$SR time spectra for a polycrystalline LBCO-1/8 
sample at ambient and at maximum applied pressure $p$ = 2.2 GPa, respectively. Below $T$ ${\approx}$ 30 K 
damped oscillations due to muon-spin precession in local magnetic fields are observed,
indicating static spin-stripe order \cite{Luke,Nachumi}. 

\begin{figure}[t!]
\includegraphics[width=0.8\linewidth]{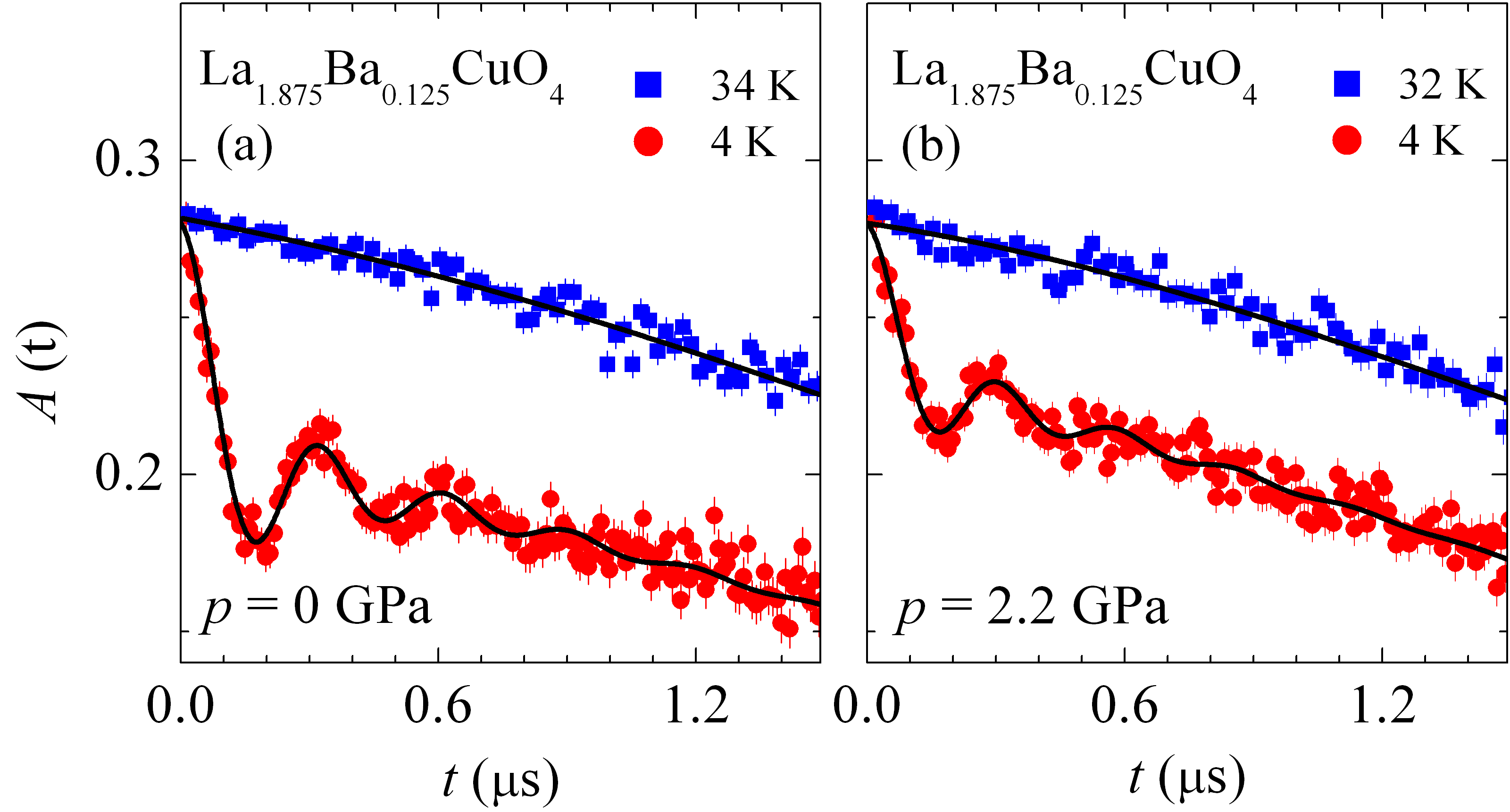}
\vspace{-0.5cm}
\caption{ (Color online) ZF ${\mu}$SR signal $A$(t) of LBCO-1/8
measured at $p$ = 0 GPa (a), and 2.2 GPa (b), recorded for two different temperatures: 
$T$ = 4 K (circles) and $T$ = 32 K (squares). 
The solid lines represent fits to the data by means of Eq.~(1).}
\label{fig1}
\end{figure}

\begin{figure}[t!]
\includegraphics[width=0.6\linewidth]{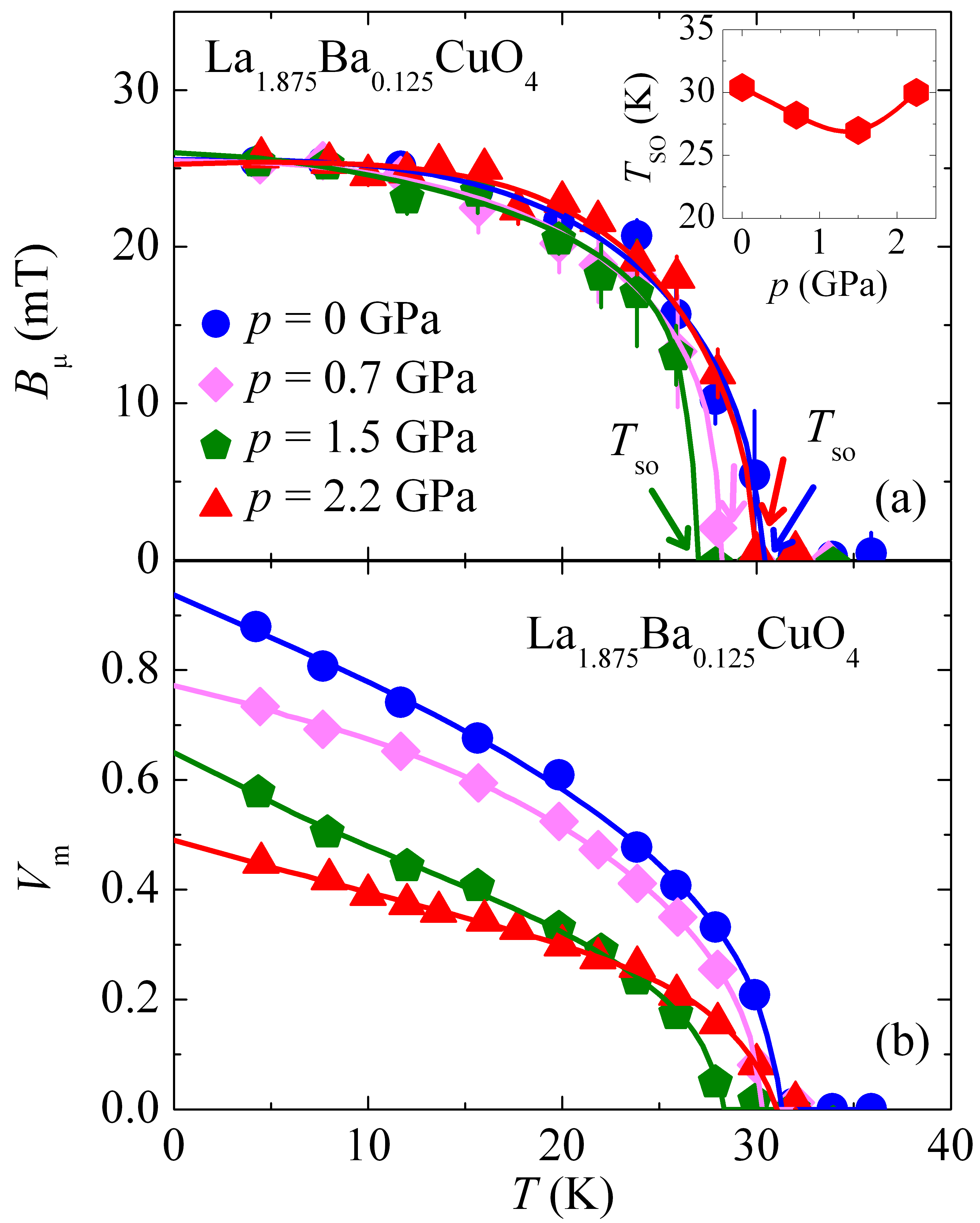}
\vspace{-0.3cm}
\caption{ (Color online) (a) Temperature dependence of the average internal 
magnetic field $B_{\rm \mu}$ at the muon site of LBCO-1/8  
recorded at various applied pressures. The solid lines represent fits of 
the data to the power law described in the text. The arrows mark the transition temperatures
for the static spin-stripe order $T_{\rm so}$. The inset shows 
$T_{\rm so}$ as a function of pressure $p$. 
(b) The temperature dependence of the magnetic 
volume fraction $V_{\rm m}$ in LBCO-1/8 at ambient and various hydrostatic pressures.
The solid lines are fits of the data to a similar empirical power law as used for $B_{\rm \mu}$($T$) in (a).}  
\label{fig1}
\end{figure}

A substantial fraction of the ${\mu}$SR asymmetry signal originates 
from muons stopping in the MP35N pressure cell surrounding the sample. 
Therefore, the ${\mu}$SR data in the whole temperature range were analyzed by
decomposing the signal into a contribution of the sample and a contribution of the pressure cell:
\begin{equation}
A(t)=A_S(0)P_S(t)+A_{PC}(0)P_{PC}(t),
\end{equation}
where $A_{S}$(0) and $A_{PC}$(0) are the initial asymmetries and $P_{S}$(t) and $P_{PC}$(t) 
are the muon-spin polarizations belonging to the sample and the pressure cell, respectively.
The pressure cell signal was analyzed by a damped Kubo-Toyabe function \cite{Maisuradze}.
The response of the sample consists of a magnetic and a nonmagnetic contribution: 
\begin{equation}
P_S(t)=V_{m}\Bigg[{\frac{2}{3}e^{-\lambda_{T}t}J_0(\gamma_{\mu}B_{\mu}t)}+\frac{1}{3}e^{-\lambda_{L}t}\Bigg]
+(1-V_{m})e^{-\lambda_{nm}t}.
\label{eq1}
\end{equation}
Here, $V_{\rm m}$ denotes the relative volume of the magnetic fraction, and $\gamma_{\mu}/(2{\pi}) \simeq 135.5$~MHz/T 
is the muon gyromagnetic ratio. 
$B_{\mu}$ is the average internal magnetic field at the muon site. ${\lambda_T}$ and ${\lambda_L}$
are the depolarization rates representing the transversal and the longitudinal 
relaxing components of the magnetic parts of the sample.
$J_{0}$ is the zeroth-order Bessel function of the first kind.
This is characteristic for an incommensurate spin density wave 
and has been observed in cuprates with static spin stripe order \cite{Nachumi}.
${\lambda_{nm}}$ is the relaxation rate of the nonmagnetic part of the sample.
The total initial assymetry $A_{\rm tot}$ = $A_{\rm S}$(0) + $A_{\rm PC}$(0) ${\simeq}$ 0.285 is a temperature independent constant. 
A typical fraction of muons stopped in the sample was $A_{\rm S}$(0)/$A_{\rm tot}$ ${\simeq}$ 0.50(5)
which was assumed to be temperature independent in the analysis. 

The temperature dependence of $B_{\rm \mu}$ for different pressures is shown in Fig.~3a. 
The solid curves in Fig.~3a are fits of the data to the power law 
$B_{\mu}$($T$) = $B_{\mu}$(0)[1-($T/T_{\rm so}$)$^{\gamma}$]$^{\delta}$,
where $B_{\rm \mu}$(0) is the zero-temperature value of $B_{\mu}$. ${\gamma}$ and ${\delta}$ are phenomenological exponents. 
The values of the spin ordering temperature $T_{\rm so}$ ${\simeq}$ 30 K and $B_{\rm \mu}$(0) ${\simeq}$ 25 mT  
at ambient pressure are in good agreement with the values of a previous ${\mu}$SR study \cite{Satoh,Nachumi}. 
As evident from Fig.~3a the internal magnetic field $B_{\mu}$(0) is almost pressure independent.
This indicates that the ordered magnetic moment of the static stripe phase does not depend on applied pressure.
Also $T_{\rm so}$ changes only slightly with pressure as shown in the inset of Fig.~3a. 
In the pressure range of $p$ = 0 - 2.2 GPa, $T_{\rm so}$($p$) varies only between 30 and 27 K with a shallow minimum at 
$p$ ${\simeq}$ 1.5 GPa. 

It is important to note that both the LTT and LTO structural phase transition are suppressed at $p_{\rm c}$ = 1.85 GPa \cite{Hucker}. 
Therefore, the present results demonstrate that the spin order due to static stripes still exists at 
$p$ = 2.2 GPa, where the LTT phase is already suppressed.  Recent high pressure x-ray diffraction experiments showed that 
also the charge order of the stripe phase survives above $p_{\rm c}$ in LBCO-1/8 \cite{Hucker}. 
Combining these results, one may conclude that both charge and spin order, and consequently the static stripe phase itself,
still exist at pressures where the LTT phase is suppressed.  

Here the question arises: What is the effect of pressure on the stripe order in LBCO-1/8?
In agreement with the previous low-pressure ${\mu}$SR results \cite{Satoh}, it was found that it is the magnetic volume fraction $V_{\rm m}$ which is significantly suppressed by pressure. 
${\mu}$SR can determine the ordered volume fraction and is thus a particularly powerful
tool to study inhomogeneous magnetism in materials. 
Figure ~3b shows the temperature dependence of $V_{\rm m}$ at various pressures. 
$V_{\rm m}$ increases progressively below $T_{\rm so}$ with decreasing temperature and acquires
nearly 100 {\%} at ambient pressure at the base temperature \cite{Luke}. An important result is that 
at low temperature $V_{\rm m}$ significantly decreases with increasing pressure (see Fig.~3b). 
This means that with increasing pressure an increasingly large part of the sample remains in the nonmagnetic state 
down to the lowest temperatures. 

\begin{figure}[t!]
\includegraphics[width=0.6\linewidth]{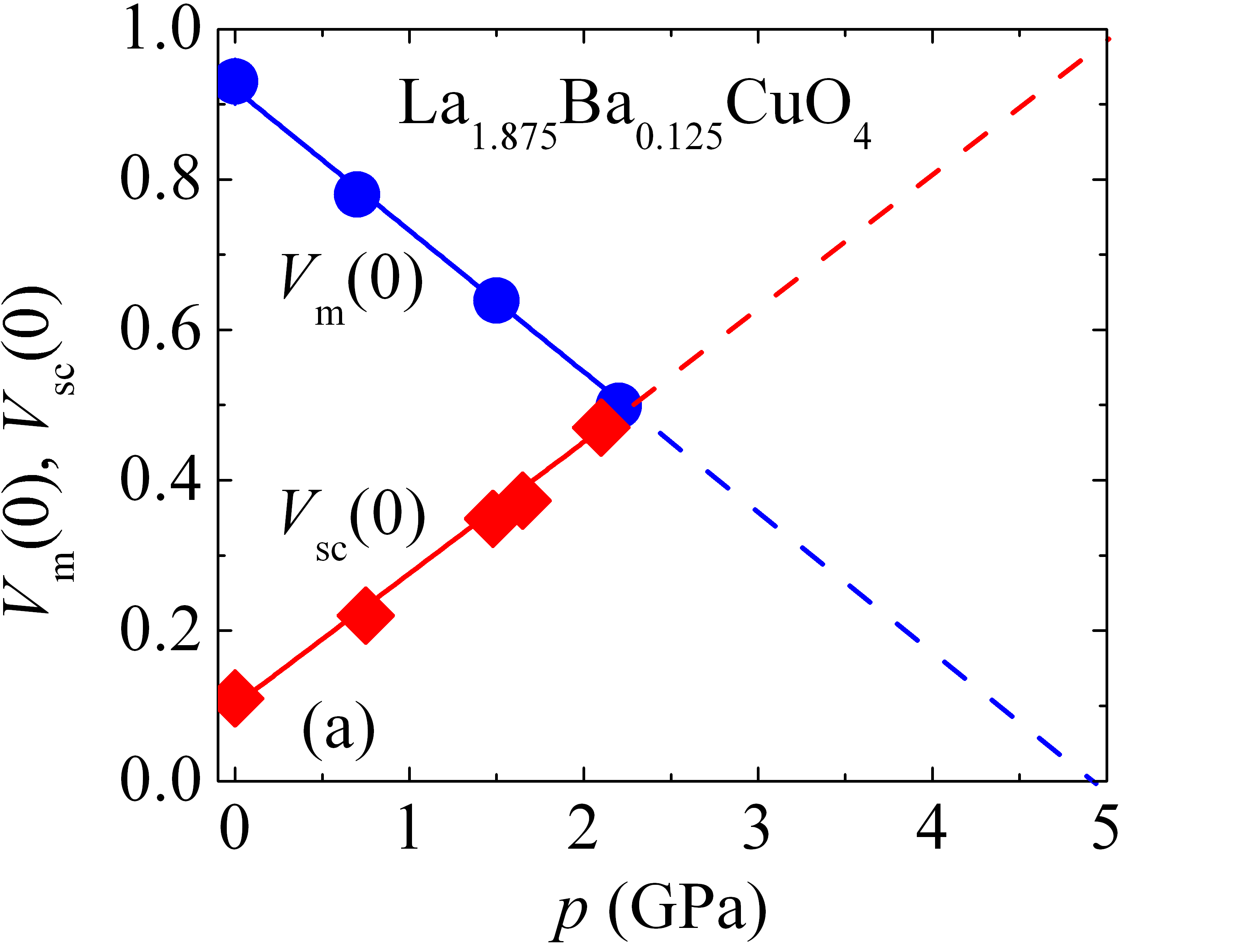}
\includegraphics[width=0.6\linewidth]{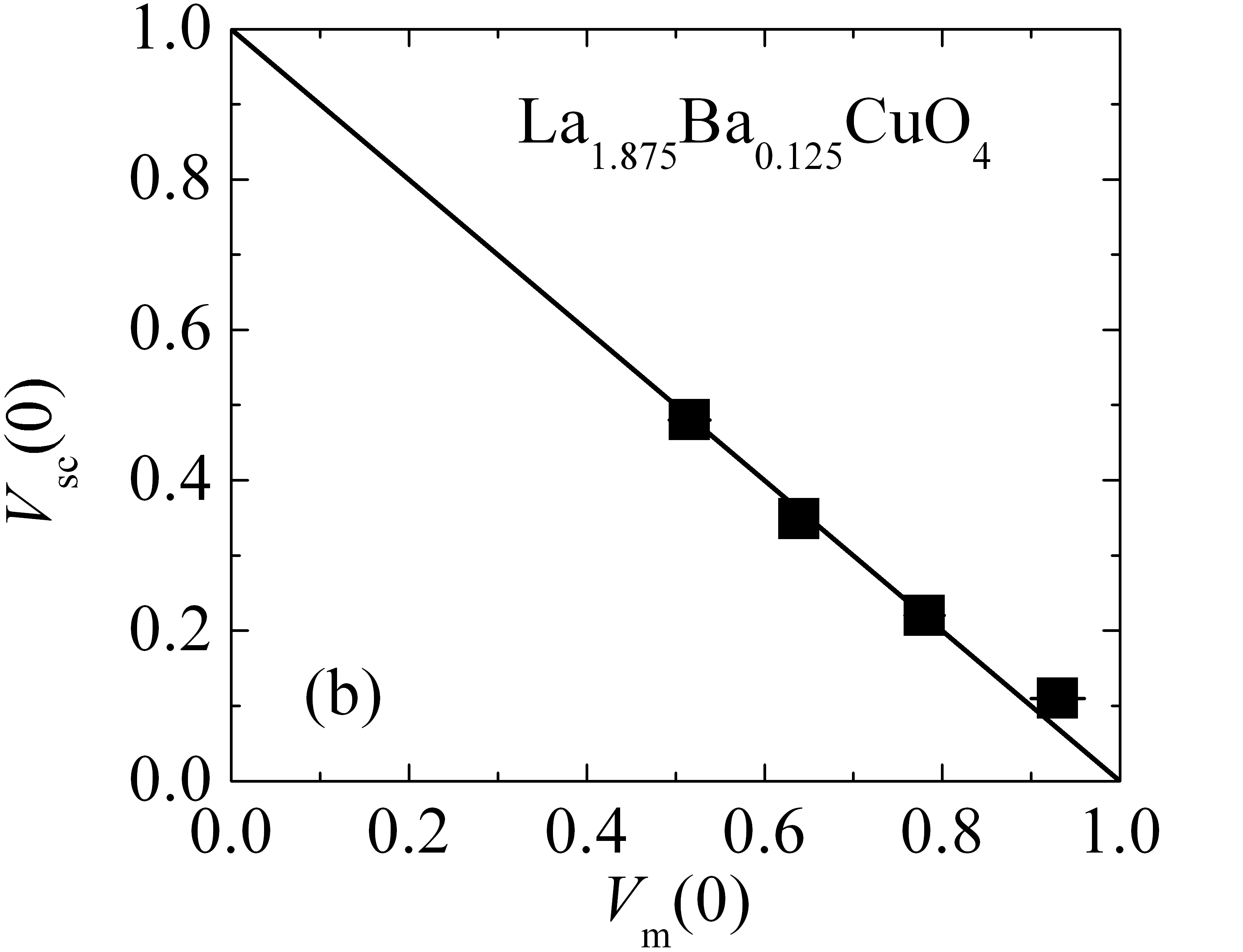}
\vspace{-0.3cm}
\caption{ (Color online) (a) The pressure dependence of the zero-temperature limit of the magnetic and
the SC volume fractions, $V_{\rm m}$(0) and $V_{\rm sc}$(0), respectively, of LBCO-1/8.
Solid lines are linear fits to the data.  
(b) $V_{\rm sc}$(0) vs. $V_{\rm m}$(0). The solid straight line is drawn 
between a hypothetical situation of a fully magnetic ($V_{\rm m}$(0) = 1) and 
a fully SC state ($V_{\rm sc}$(0) = 1).}
\label{fig1}
\end{figure}

In order to compare the influence of pressure on the SC and magnetic properties of
LBCO-1/8, the pressure dependences of the zero-temperature limit of the magnetic volume fraction $V_{\rm m}$(0) and
the SC volume fraction $V_{\rm sc}$(0) = -${\chi}_{\rm ZFC}$(0) \cite{comment} are plotted in Fig.~4a.
Note that $V_{\rm m}$(0) linearly decreases with pressure to
approximately 50 {\%} at $p$ = 2.2 GPa. A linear extrapolation of $V_{\rm m}$(0) 
to higher pressures shows that the magnetic volume fraction should be completely suppressed 
at $p$ ${\approx}$ 5 GPa. It would be interesting to check this prediction
at higher pressures by either ${\mu}$SR or neutron-scattering experiments.
It is evident from Fig.~4a that the decrease of $V_{\rm m}$(0) is followed by an increase
of the SC volume fraction $V_{\rm sc}$(0). In Fig.~4b we plot $V_{\rm sc}$(0)  
as a function of $V_{\rm m}$(0).
The solid straight line is drawn between a hypothetical situation of a fully magnetic ($V_{\rm m}$(0) = 1) and 
a fully SC state ($V_{\rm sc}$(0) = 1). 
Remarkably, the experimental data lie on this solid straight line. 
Thus, the sum of the SC and magnetic volume fractions is constant and is close to one. 
This strongly suggests that superconductivity does not exist in those regions where static magnetism
is present. Thus, superconductivity most likely develops in those areas of the sample which are nonmagnetic down to the lowest
temperatures. The latter implies that in LBCO-1/8 magnetism and superconductivity
are competing order parameters. It is interesting to note that a similar scaling was found 
between the superfluid density and the magnetic volume fraction in the related compound
La$_{1.85-y}$Eu$_{y}$Sr$_{0.15}$CuO$_{4}$ \cite{Kojima}. The tuning
of the magnetic and SC properties was realized by rare-earth doping.

To summarize, magnetism and superconductivity was studied
in LBCO-1/8 by means of magnetization and ${\mu}$SR experiments as
a function of pressure up to $p$ ${\simeq}$ 2.2 GPa. 
It was demonstrated that the static spin-stripe order still exist at pressures,
where the long-range LTT structure is suppressed. This suggests that the long-range LTT phase is not essential for the 
existence of stripe order.
An unusual interplay between spin order and bulk superconductivity was also observed. With increasing pressure
the spin order temperature and the size of the ordered moment are not changing significantly. 
However, application of hydrostatic pressure 
leads to a remarkable decrease of the magnetic volume fraction $V_{\rm m}$(0). Simultaneously, an increase of the 
SC volume fraction $V_{\rm sc}$(0) occurs. Furthermore, it was found that $V_{\rm m}$(0) and 
$V_{\rm sc}$(0) at all $p$ are linearly correlated: $V_{\rm m}$(0) + $V_{\rm sc}$(0) ${\simeq}$ 1. 
This is an important new result, indicating that the
magnetic fraction in the sample is  directly converted to the SC fraction with increasing pressure.
The mechanism of this transformation, however, is not clear yet and requires further studies. 
The present results provide evidence for a competition 
between bulk superconductivity and static magnetic order in the stripe phase of LBCO-1/8, 
and that static stripe order and bulk superconductivity occur in mutually exclusive spatial regions. 
Our findings suggest that a pressure of about 5 GPa would be sufficient to completely suppress the static stripe phase
and restore bulk superconductivity in LBCO-1/8.

The ${\mu}$SR experiments were performed at the Swiss Muon Source, Paul Scherrer Institute (PSI),
Villigen, Switzerland. This work was supported by the Swiss National Science Foundation, the NCCR MaNEP, 
the SCOPES grant No. IZ73Z0-128242, and the Georgian National Science Foundation grant RNSF/AR/10-16.



\begin{thebibliography}{22}
\bibitem{comment} As mentioned in the text we measured only ${\chi}_{\rm FC}$ under high pressure. 
The difference ${\chi}_{\rm ZFC}$(0)-${\chi}_{\rm FC}$(0) was obtained
from the measurements at ambient pressure (see Fig.~1a). By assuming that this difference does not change 
with pressure, the value of ${\chi}_{\rm ZFC}$ at $T$ = 0 K was determined at different applied pressures.
\end{thebibliography}
\end{document}